\documentclass{iau}
\usepackage{graphicx}

\title[Magnetic Activity of Pre-Main Sequence Stars near the Stellar-Substellar Boundary] 
{Magnetic Activity of Pre-main Sequence Stars near the Stellar-Substellar Boundary}

\author[D. A. Principe, J. H. Kastner, \& D. Rodriguez]   
{David A. Principe$^{1,}$$^{2}$, Joel. H. Kastner$^3$
 \and David Rodriguez$^4$}

\affiliation{$^1$N{\' u}cleo de Astronom{\' i}a, Facultad de Ingenier{\' i}a, Universidad Diego Portales, Santiago, Chile\\email: {\tt daveprincipe1@gmail.com}\\
$^2$Millennium Nucleus Protoplanetary Disks, Chile\\
$^3$Center for Imaging Science, School of Physics \& Astronomy, and Laboratory for Multiwavelength Astrophysics, Rochester Institute of Technology, Rochester, NY 14623, USA \\
$^4$Departamento de Astronom{\' i}a, Universidad de Chile, Casilla 36-D, Santiago, Chile\\}

\pubyear{2015}
\volume{314}  
\pagerange{119--126}
\jname{Young Stars \& Planets Near the Sun}
\editors{J. H. Kastner, B. Stelzer, \& S. A. Metchev, eds.}
\begin{document}

\maketitle

\begin{abstract}
 X-ray observations of pre-main sequence (pre-MS) stars of M-type probe coronal emission and offer a means to investigate magnetic activity at the stellar-substellar boundary. Recent observations of main sequence (MS) stars at this boundary display a decrease in fractional X-ray luminosity ($L_{X}$/$L_{bol}$) by almost two orders of magnitude for spectral types M7 and later.  We investigate magnetic activity and search for a decrease in X-ray emission in the pre-MS progenitors of these MS stars.  We present XMM-Newton X-ray observations and preliminary results for $\sim$10 nearby (30-70 pc), very low mass pre-MS stars in the relatively unexplored age range of 10-30 Myr.  We compare the fractional X-ray luminosities of these 10-30 Myr old stars to younger (1-3 Myr) pre-MS brown dwarfs and find no dependence on spectral type or age suggesting that X-ray activity declines at an age later than $\sim$30 Myr in these very low-mass stars.

\keywords{\vspace{-2mm}Stars: formation, Stars: magnetic field, Stars: late-type, X-rays: stars. \vspace{-2mm}}
\end{abstract}

\firstsection 

\vspace{-3mm}
\section{Introduction}

 The early evolution of magnetic activity in very low mass pre-MS stars --stars of mid-M-type, which lie near the H-burning limit of 0.08 $M_{\odot}$-- is very poorly understood. Yet understanding their pre-MS evolution is crucial for determining the emerging differences between very low-mass MS stars and brown dwarfs. X-ray emission offers a means to indirectly probe the effects of internal and surface magnetic activity in both pre-MS and MS stars alike (\cite[Vidotto et al. 2014]{Vidotto2014}).  Pre-MS and MS M-type stars are magnetically active and thus can be bright X-ray sources, as indicated by their high values of ($L_{X}$/$L_{bol} \sim 10^{-3}$).  However, observations of nearby late M-type MS stars suggest that stars of $\sim$M7 and later appear to be under luminous in X-rays (e.g., $L_{X}$/$L_{bol}$ $\sim$ $10^{-5}$; \cite[Berger et al. 2010]{Berger2010}).  The narrow range of spectral types where these M stars become X-ray under luminous is roughly the same spectral type where a transition to predominantly neutral atmospheres occurs (\cite[Mohanty et al. 2002]{Mohanty2002}). \cite[Berger (2006)]{Berger2006} concluded that the decrease in X-ray activity (as well as H$\alpha$) toward late M-types is related to changes in magnetic field configuration or the decreasing ionization fractions in the atmospheres of these stars.

By determining the age at which this dramatic decrease in X-ray activity occurs for M-type stars, we can gain insight into the early pre-MS stellar evolution of such stars which lie at the low-mass-star/brown dwarf (H-burning) boundary.  A recent survey combining GALEX, 2MASS, WISE and catalog proper motions have revealed a population of nearby late-M-type stars in the 10-30 Myr age range (\cite[Rodriguez et al. 2013]{Rodriguez2013}) where X-ray activity of such stars has remained, until now, essentially unexplored.

\vspace{-5mm}
\section{Data and Preliminary Results}

We performed XMM-Newton EPIC X-ray observations of 8 nearby ($<$70 pc) $\sim$M5 members of the Tuc-Hor and $\beta$ Pic moving groups (ages $\sim$ 30 Myr and 12 Myr, respectively; Rodriguez et al. 2013 and ref. therein).  Stellar evolution models (\cite[D'Antona \& Mazzitelli 1997]{Dantona1997}) suggest pre-MS stars of this age and spectral type will evolve to become MS $\sim$M7 (i.e., they may be progenitors of the under luminous MS M7 stars). Standard one and two temperature thermal plasma models were fit to the data to determine spectral parameters such as plasma temperature and $L_{X}$.  Bolometric luminosities for each of our sources was estimated using their J band flux and the intrinsic colors of 5-30 Myr stars from \cite[Pecaut \& Mamajek (2013)]{Pecaut2013}.  The fractional X-ray luminosity for each source is shown in Fig. 1 and compared to pre-MS stars of similar spectral type in younger (e.g 1-3 Myr) star-forming regions.

\vspace{-6mm}
\section{Conclusions}

We find no trend of decreasing fractional X-ray luminosity with age in these 10-30 Myr $\sim$M5 stars (Fig. 1).  If MS stars of $\sim$M7 and later are under luminous in X-rays, these preliminary results suggest that either X-ray activity decreases at ages later than $\sim$30 Myr or that our sample of pre-MS stars is unusually magnetically active.  The latter scenario may be more likely due to the fact that the sample from which we chose our sources required a GALEX UV detection (i.e., only M stars that were UV bright).  UV emission is also an indicator of magnetic activity and thus our sample might be biased towards magnetically active (i.e., X-ray bright) pre-MS stars.  More X-ray observations of both UV bright and UV faint mid-to-late M-type pre-MS stars are required to explore the age at which X-ray activity may diminish. A more detailed analysis of these data will be presented in Principe et al. (2015), in prep.

\begin{figure}[b]
\begin{center}
\vspace{-4.5mm}
\includegraphics[width=2.87in]{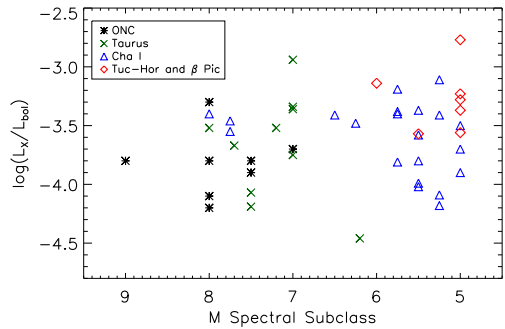}
 \caption{Fractional X-ray luminosity as a function of M spectral subclass for $\sim$1-3 Myr old pre-MS stars in the Orion Nebula Cluster (asterisk), Taurus (cross), Cha I (triangle) and preliminary results of $\sim$10-30 Myr old Tuc-Hor and $\beta$ Pic moving group members (diamonds). Figure originally from \cite[Stelzer \& Micela (2007)]{Stelzer2007} and modified to include preliminary results. }
\label{fig:grains}
\end{center}
\end{figure}

\end{document}